\documentclass[runningheads]{llncs}
\usepackage[T1]{fontenc}
\usepackage{graphicx}
\usepackage{threeparttable}
\usepackage{caption}
\usepackage{amsmath}
\usepackage{amssymb}
\usepackage{booktabs}
\usepackage{adjustbox}
\usepackage{multirow}
\usepackage{diagbox}
\usepackage{float}
\usepackage{colortbl}
\usepackage{xcolor}
\usepackage{array}
\usepackage[pagebackref,breaklinks,colorlinks]{hyperref}
\usepackage[linesnumbered,ruled]{algorithm2e}
\usepackage{tikz}
 
\usepackage{multirow}
\begin{document}
%
% \title{Contribution Title\thanks{Supported by organization x.}}
% \title{DiffuseIR: Isotropic Reconstruction of 3D Microscopic Images Using Diffusion Models}

\title{DiffuseIR: Diffusion Models For Isotropic Reconstruction of 3D Microscopic Images}

\author{
Mingjie Pan\textsuperscript{\rm 1}\thanks{Equal contribution},
Yulu Gan\textsuperscript{\rm 1*},
Fangxu Zhou\textsuperscript{\rm 1}, \\
Jiaming Liu\textsuperscript{\rm 1},
Aimin Wang\textsuperscript{\rm 1},
Shanghang Zhang\textsuperscript{\rm 1},
Dawei Li\textsuperscript{\rm 1},   \\
\textsuperscript{\rm 1}Peking University\\ 
}

\maketitle              % typeset the header of the contribution
\begin{abstract}

Three-dimensional microscopy is often limited by anisotropic spatial resolution, resulting in lower axial resolution than lateral resolution. Current State-of-The-Art (SoTA) isotropic reconstruction methods utilizing deep neural networks can achieve impressive super-resolution performance in fixed imaging settings. However, their generality in practical use is limited by degraded performance caused by artifacts and blurring when facing unseen anisotropic factors. To address these issues, we propose DiffuseIR, an unsupervised method for isotropic reconstruction based on diffusion models. First, we pre-train a diffusion model to learn the structural distribution of biological tissue from lateral microscopic images, resulting in generating naturally high-resolution images. Then we use low-axial-resolution microscopy images to condition the generation process of the diffusion model and generate high-axial-resolution reconstruction results. Since the diffusion model learns the universal structural distribution of biological tissues, which is independent of the axial resolution, DiffuseIR can reconstruct authentic images with unseen low-axial resolutions into a high-axial resolution without requiring re-training. The proposed DiffuseIR achieves SoTA performance in experiments on EM data and can even compete with supervised methods.

\end{abstract}
\section{Introduction}
% \textcolor{red}{[Background]}
Three-dimensional (3D) microscopy imaging is crucial in revealing biological information from the nanoscale to the microscale. Isotropic high resolution across all dimensions is desirable for visualizing and analyzing biological structures. 
However, most three-dimensional imaging techniques often have lower axial (z) resolution than lateral (xy) resolution, due to physical slicing interval limitation (serial section transmission electron microscopy, automated tape-collecting ultra-microtome scanning electron microscopy, etc.) \cite{ShawnMikula2016ProgressTM} or time-saving consideration (focused ion beam scanning electron microscopy, fluorescence microscopy, etc.) \cite{KennethJHayworth2015UltrastructurallyST,YichenWu2019ThreedimensionalVR,PeterJVerveer2007HighresolutionTI,TinaSchrdel2013Brainwide3I}. 
Therefore, effective isotropic super-resolution algorithms are critical for high-quality 3D image reconstructions, such as electron microscopy and fluorescence microscopy.
% 通常，医生使用插值作为恢复轴向分辨率的处理方法，但是会引起axial模糊和细节丢失，影响后续分析。

% \textcolor[rgb]{1,0,0}{(Related work)}
% Recently, methods based on deep learning have made significant progress in natural image restoration and biomedical image analysis. 
Recently, deep learning methods have made significant progress in image analysis \cite{LarissaHeinrich2017DeepLF,YangSong2021SolvingIP,sr1,sr2}. 
To address the isotropic reconstruction problem, \cite{LarissaHeinrich2017DeepLF} employs isotropic EM images to generate HR-LR pairs at axial and train a super-resolution model in a supervised manner, demonstrating the feasibility of inferring HR structures from LR images. \cite{MartinWeigert2017IsotropicRO,MartinWeigert2018ContentAwareIR} use 3D point spread function (PSF) as a prior for self-supervised super-resolution. 
However, isotropic high-resolution images or 3D point spread function (PSF) physical priors are difficult to obtain in practical settings, thus limiting these algorithms.
% 完全无监督Baselines
% In order to get rid of the dependence on isotropic data or physical prior, some unsupervised methods \cite{ShiyuDeng2020IsotropicRO,HyoungjunPark2021DeepLE} skillfully use the periodic consistent generation network (cycleGAN), learning from an unpaired matching between high-resolution 2D images in the lateral plane and low-resolution 2D images in the axial plane. These methods can train high-quality super-resolution models without isotropic data while achieving impressive performance. 
Some methods like \cite{ShiyuDeng2020IsotropicRO,HyoungjunPark2021DeepLE} have skillfully used cycleGAN \cite{JunYanZhu2017UnpairedIT} architecture to train axial super-resolution models without depending on isotropic data or physical priors. They learn from unpaired matching between high-resolution 2D slices in the lateral plane and low-resolution 2D slices in the axial plane, achieving impressive performance.
However, these methods train models in fixed imaging settings and suffer from degraded performance caused by artifacts and blurring when facing unseen anisotropic factors. This limits their generality in practice \cite{VGonzlezRuiz2022OpticalFD}. In conclusion, a more robust paradigm needs to be proposed.
% \textcolor[rgb]{1,0,0}{(Our method)}
Recently, with the success of the diffusion model in the image generation field \cite{XuanSu2023DualDI,AlexNichol2021ImprovedDD,PrafullaDhariwal2021DiffusionMB,AndreasLugmayr2023RePaintIU,BahjatKawar2023DenoisingDR}, researchers applied the diffusion model to various medical image generation tasks and achieved impressive results \cite{YangSong2021SolvingIP,ChengPeng2023TowardsPA,HyungjinChung2023ScorebasedDM,BoahKim2023DiffusionDM,MuzafferOzbey2022UnsupervisedMI}. Inspired by these works, we attempt to introduce diffusion models to address the isotropic reconstruction problem. 

This paper proposes DiffuseIR, an unsupervised method based on diffusion models, to address the isotropic reconstruction problem. Unlike existing methods, DiffuseIR does not train a specific super-resolution model from low-axial-resolution to high-axial-resolution. Instead, we pre-train a diffusion model $\epsilon_\theta$ to learn the structural distribution $p_\theta(X_{lat})$ of biological tissue from lateral microscopic images $X_{lat}$, which resolution is naturally high. Then, as shown in Fig. \ref{framework}, we propose a Sparse Spatial Condition Sampling (SSCS) to condition the reverse-diffusion process of $\epsilon_\theta$. SSCS extracts sparse structure context from low-axial-resolution slice $x_{axi}$ and generate reconstruction result $x_0 \sim p_\theta(X_{lat}|x_{axi})$. Since $\epsilon_\theta$ learns the universal structural distribution $p_\theta$, which is independent of the axial resolution, DiffuseIR can leverage the flexibility of SSCS to reconstruct authentic images with unseen anisotropic factors without requiring re-training. To further improve the quality of reconstruction, we propose a Refine-in-loop strategy to enhance the authenticity of image details with fewer sampling steps.
% Note that, we are the first to introduce diffusion models for isotropic reconstruction, which address the problem in an unsupervised and robust way.

% \textcolor[rgb]{1,0,0}{(Contribution)}
To sum up, our contributions are as follows: 

(1) We are the first to introduce diffusion models to isotropic reconstruction and propose DiffuseIR. Benefiting from the flexibility of SSCS, DiffuseIR is naturally robust to unseen anisotropic spatial resolutions.
(2) We propose a Refine-in-loop strategy, which maintains performance with fewer sampling steps and better preserves the authenticity of the reconstructed image details.
(3) We perform extensive experiments on EM data with different imaging settings and achieve SOTA performance. Our unsupervised method is competitive with supervised methods and has much stronger robustness.

% Our method is simple and easy to use. A model can be directly applied to data with different axial resolutions and has strong generalization. This feature is not available in previous methods and can be widely used in practical scenarios.

% Please note that the first paragraph of a section or subsection is not indented. The first paragraph that follows a table, figure, equation, etc. does not need an indent, either.

% Subsequent paragraphs, however, are indented. badi

\section{Methodology}

\begin{figure}[t]
\centering
\includegraphics[width=1.0\linewidth]{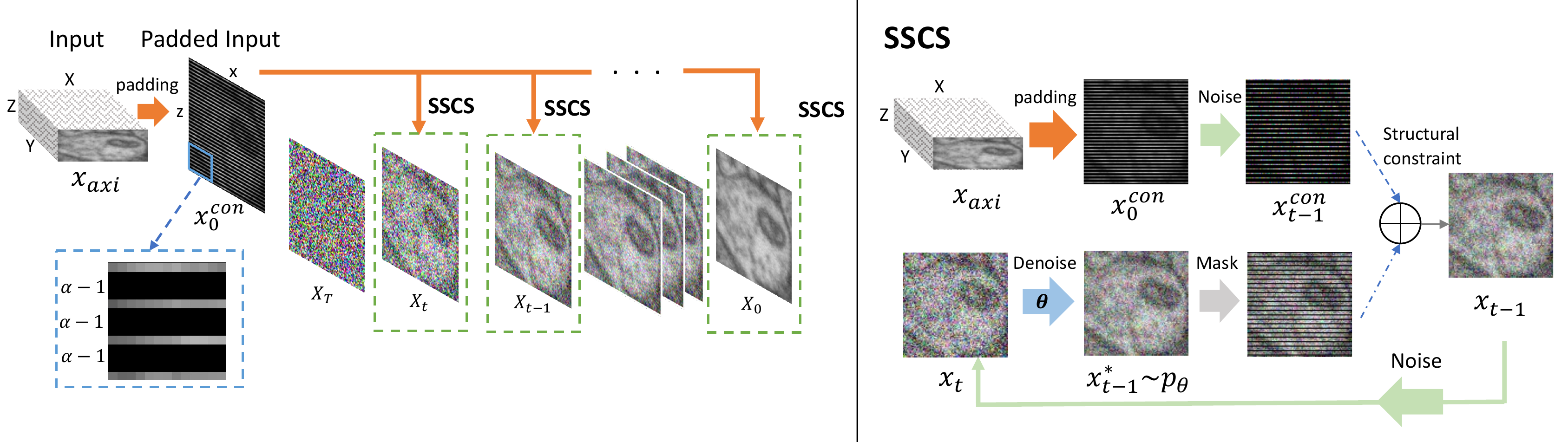}     
\caption{\label{fig:1}\textbf{Method Pipeline.} 
% DiffuseIR progressively conditions the denoising process with Sparse Spatial Condition Sampling (SSCS). 
DiffuseIR progressively conditions the denoising process with SSCS. 
For SSCS, we perform intra-row padding on input $X_{lat}$ using the anisotropy factor $\alpha$ to obtain spatially aligned structural context, which is then merged with the diffusion model's output. Iterative SSCS refines reconstruction.
}

\label{framework}
\end{figure}
As shown in Fig. \ref{framework}, DiffuseIR address isotropic reconstruction by progressively conditions the denoising process of a pre-trained diffusion model $\epsilon_\theta$. Our method consists of three parts: DDPM pre-train, Sparse Spatial Condition Sampling and Refine-in-loop strategy.

\subsubsection{DDPM Pretrain on lateral}
% \subsection{Learning prior knowledge by DDPM}
Our method differs from existing approaches that directly train super-resolution models. Instead, we pre-train a diffusion model to learn the distribution of high-resolution images at lateral, avoiding being limited to a specific axial resolution.
Diffusion models \cite{JonathanHo2020DenoisingDP,AlexNichol2021ImprovedDD} employ a Markov Chain diffusion process to transform a clean image $x_0$ into a series of progressively noisier images during the forward process. This process can be simplified as:
\begin{equation}
\begin{aligned}
\label{add_noise_1}
q(x_t|x_0) = N(x_t;\sqrt{\overline{\alpha}_t}x_0,(1-\overline{\alpha}_t)I),
\end{aligned}
\end{equation}
% with $\alpha_t=1-\beta_t$ and $\overline{\alpha}_t=\prod^{t}_{i=1}\alpha_i$.
where $\overline{\alpha}_t$ controls the scale of noises.
During inference, the model $\epsilon_\theta$ predicts $x_{t-1}$ from $x_{t}$. A U-Net $\epsilon_\theta$ is trained for denoising process $p_\theta$, which gradually reverses the diffusion process. This denoising process can be represented as:
\begin{equation}
\begin{aligned}
\label{de_diffusion}
p_\theta(x_{t-1}|x_{t}) = N(x_{t-1};\epsilon_\theta(x_t,t),\sigma_t^2I),
\end{aligned}
\end{equation}
% \begin{equation}
% \begin{aligned}
% \label{de_diffusion_2}
% \sigma_t^2=\overline{\beta_t}=\frac{1-\overline{\alpha}_{t-1}}{1-\overline{\alpha}_t}
% \end{aligned}
% \end{equation}
% Note that $\sigma_t^2=\overline{\beta_t}=\frac{1-\overline{\alpha}_{t-1}}{1-\overline{\alpha}_t}$.
During training, we use 2D lateral slices, which is natural high-resolution to optimize $\epsilon_\theta$ by mean-matching the noisy image obtained in Eq. \ref{add_noise_1} using the MSE loss \cite{JonathanHo2020DenoisingDP}.
%We follow \cite{AlexNichol2021ImprovedDD} and train $\epsilon_\theta$ on high-resolution(HR) 3D microscopes slices. 
% We train $\epsilon_\theta$ on high-resolution(HR) 3D microscopes slices. Note that only $X_{lat}$ (HR slices at lateral plane) were used to build the training set, so the model training neither rely on HR ground truth at axial nor require any physical priors. Instead, a generative task is used to capture the distribution of $X_{lat}$.
Only HR slices at lateral plane $X_{lat}$ were used for training, so the training process is unsupervised and independent of the specific axial resolution. 
% Benefiting from the power generation ability of diffusion models, $\epsilon_\theta$ can learn rich and detailed semantic information of biological tissue.
 So that $\epsilon_\theta$ learns the universal structural distribution of biological tissues and can generate realistic HR images following $p_\theta(X_{lat})$.

% so the training process neither rely on HR ground truth at axial nor require physical priors. Instead, we use the diffusion model to learn the universal structural distribution of $X_{lat}$, which is independent of the axial resolution.
% learns the universal structural distribution of biological tissues, which is independent of the axial resolution,
% $x_{HR}$, and then we can transform noise and generate images that follow the distribution of $x_{HR}$.

\begin{algorithm}[htb]
\DontPrintSemicolon
  \SetAlgoLined
  \KwIn {axial slice $x_{axi}$, anisotropic
factor $\alpha$, refine-in-loop counts $K$}
  % \KwOut {$x_0$}
  % initialization\;
  $x^{con}_0, M \gets padding(x_{axi}, \alpha)$\;
  \For{$t=T,...,1$}{
    $x^{con}_{t-1} \sim N(\sqrt{\overline{\alpha}_t}x^{con}_0,(1-\overline{\alpha}_t)I)$ \;
    \For {$i=1,...,K$}{
        $x_{t-1}^{*} \sim N(x_{t-1};\epsilon_\theta(x_t,t),\sigma_t^2I)$ \;
        % $x_{t-1}^{*} = \sqrt{\overline{\alpha}_{t-1}}(\frac{x_t-\sqrt{1-\overline{\alpha}_t}\epsilon_\theta(x_t,t)}{\sqrt{\overline{\alpha}_t}}) + \sqrt{1-\overline{\alpha}_{t-1}-\sigma_t^2}\epsilon_\theta(x_t,t) + \sigma_tz_t$ \;
        % N(x_{t-1};,\sigma_t^2I)$ \;
        
        $x_{t-1} = M*x^{con}_{t-1}+(1-M)*x_{t-1}^{*}$ \;
        \If{$t>1$ and $i<K$}{
            $x_t \sim N(\sqrt{1-\beta_t}x_{t-1},\beta_tI)$\;
            % $x_t \sim N(\sqrt{\overline{\alpha_t}x_{t-1},\beta_tI)$\;
        }
        
    }
  }
  return $x_0$
  \caption{Isotropic reconstruction using basic DiffuseIR}
\end{algorithm}

\subsubsection{Sparse Spatial Condition Sampling on axial} 

% 我们需要模型输出的结果不仅遵循HR数据的分布，还遵循输入图像的内容。因此，我们提出了SCS，对DDPM的sampling过程添加约束以得到对特定LR图像的重建结果。
We propose Sparse Spatial Condition Sampling (SSCS) to condition the generation process of $\epsilon_\theta$ and generate high-axial-resolution reconstruction results. SSCS substitutes every reverse-diffusion step Eq. \ref{de_diffusion}. 
We first transform the input axial LR slice $x_{axi}$ to match the lateral resolution by intra-row padding: $(\alpha-1)$ rows of zero pixels are inserted between every two rows of original pixels, where $\alpha$ is the anisotropic spatial factor. We denote $M$ as the mask for original pixels in $x^{con}_0$, while $(1-M)$ represents those empty pixels inserted. In this way, we obtain $x^{con}_0$, which reflects the sparse spatial content at axial, and further apply Eq. \ref{add_noise_1} to transform noise level: 
\begin{equation}
\begin{aligned}
\label{add_noise_2}
x^{con}_{t-1} \sim N(\sqrt{\overline{\alpha}_t}x^{con}_0,(1-\overline{\alpha}_t)I)
\end{aligned}
\end{equation}
Then, SSCS sample $x_{t-1}$ at any time step $t$, conditioned on $x^{con}_{t-1}$. The process can be described as follows:
\begin{equation}
\begin{aligned}
\label{add_noise_4}
x_{t-1} = M\odot x^{con}_{t-1} + (1-M) \odot x_{t-1}^{*})
\end{aligned}
\end{equation}
where $x_{t-1}^{*}$ is obtained by sampling from the model $\epsilon_\theta$ using Eq. \ref{de_diffusion}, with $x_t$ of the previous iteration. $x_{t-1}^{*}$ and $x_{t-1}^{con}$ are combined with $M$. 
% 通过反复的进行降噪，我们最终得到$x_0$，它既服从pretrained diffusion model学到的高分辨图像分布、又与输入的低分辩轴向切片中的语义一致，得到$x_{axi}$的各向同性重构结果。由于SSCS是parameter-free的并且解耦于模型的训练过程，DiffuseIR可以适配于任何各向异性分辨率。我们只需要根据$\alpha$来修改padding系数，而不需要像其他方法一样重新训练模型，这是一个新的特性，一定程度上解决了基于深度学习的各向同性超分方法在实际场景中的通用性问题。
% With iteratively denoising, we can eventually obtain the reconstruction result $x_0$, which not only follows the distribution $p_\theta(X_{lat})$ learned by the pretrained diffusion model but also maintains semantic consistency with the input LR axial slice.  
% Since SSCS is parameter-free and decoupled from the model training process, DiffuseIR can be adapted to various anisotropic spatial resolution. We only need to modify the padding factor according to $\alpha$, without having to retrain the model as required by other methods. 
% This is a new feature that partially addresses the generality issue of DL-based isotropic reconstruction methods in practical scenarios.
By iterative denoising, we obtain the reconstruction result $x_0$. It conforms to the distribution $p_\theta(X_{lat})$ learned by the pre-trained diffusion model and maintains semantic consistency with the input LR axial slice. 
Since SSCS is parameter-free and decoupled from the model training process, DiffuseIR can adapt to various anisotropic spatial resolutions by modifying the padding factor according to $\alpha$ while other methods require re-training. This makes DiffuseIR a more practical and versatile solution for isotropic reconstruction.

\subsubsection{Refine-in-loop Strategy}
We can directly use SSCS to generate isotropic results, but the reconstruction quality is average. The diffusion model is capable of extracting context from the sparse spatial condition. Still, we have discovered a phenomenon of texture discoordination at the mask boundaries, which reduces the reconstruction quality. 
For a certain time step $t$, the content of $x_{t-1}^*$ may be unrelated to $x_{t-1}^{con}$, resulting in disharmony in $x_{t-1}$ generated by SSCS. During the denoising of the next time step $t-1$, the model tries to repair the disharmony of $x_{t-1}$ to conform to $p_\theta$ distribution. Meanwhile, this process will introduce new inconsistency and cannot converge on its own. To overcome this problem, we propose the Refine-in-loop strategy: For $x_{t-1}$ generated by SSCS at time step $t$, we apply noise to it again and obtain a new $x_t$ and then repeat SSCS at time step $t$. Our discovery showed that this uncomplicated iterative refinement method addresses texture discoordination significantly and enhances semantic precision.

 The total number of inference steps in DiffuseIR is given by $T_{total}=T \cdot K$. As $T_{total}$ increases, it leads to a proportional increase in the computation time of our method. 
 However, larger $T_{total}$ means more computational cost. Recent works such as \cite{JiamingSong2020DenoisingDI,ChengLu2022DPMSolverAF,ChengLu2022DPMSolverFS} have accelerated the sampling process of diffusion models by reducing $T$ while maintaining quality. 
 For DiffuseIR, adjusting the sampling strategy is straightforward. Lowering $T$ and raising refinement iterations $K$ improves outcomes with a fixed $T_{total}$.
 We introduce and follow the approach presented in DDIM \cite{JiamingSong2020DenoisingDI} as an example and conducted detailed ablation experiments in Sec. \ref{ablation} to verify this. 
 Our experiments show that DiffuseIR can benefit from advances in the community and further reduce computational overhead in future work.

\section{Experiments and Discussion}
\subsubsection{Dataset and implement details.}
To evaluate the effectiveness of our method, we conducted experiments on two widely used public EM datasets, FIB-25 \cite{ShinyaTakemura2015SynapticCA} and Cremi \cite{cremiorg}. FIB-25 contains isotropic drosophila medulla connectome data obtained with FIB-SEM. We partitioned it into subvolumes of 256x256x256 as ground truth and followed \cite{LarissaHeinrich2017DeepLF} to perform average-pooling by factor $\alpha$(x2,x4,x8) along the axis to obtain downsampled anisotropic data. Cremi consists of drosophila brain data with anisotropic spatial resolution. We followed \cite{ShiyuDeng2020IsotropicRO} to generate LR images with a degradation network and conduct experiments on lateral slices. All resulting images were randomly divided into the training (70\%), validation (15\%) and test (15\%) set.
For the pre-training of the diffusion model, we follow \cite{AlexNichol2021ImprovedDD} by using U-Net with multi-head attention and the same training hyper-parameters. We use 256×256 resolution images with a batch size of 4 and train the model on 8×V100 GPUs. For our sampling setting, we set ${T, K}={25,40}$, which is a choice selected from the ablation experiments in Sec. \ref{ablation} that balances performance and speed.

\begin{table*}[t]  
    \centering 
    \footnotesize
    \setlength\tabcolsep{5pt}%调列距
    \caption{\textbf{Quantitative evaluation of DiffuseIR against baselines.} PSNR$\uparrow$ and SSIM$\uparrow$ are used as evaluation metrics. We evaluated the FIB25 and Cremi datasets, considering three anisotropic spatial resolutions, $\alpha={2,4,8}$. Unlike other baselines which train a dedicated model for each $\alpha$, our method only trains a single, generalizable model.}
    \label{main_table1}   
    %\resizebox{\linewidth}{!}{
    %    \scriptsize %字体
        \begin{tabular}{cccccccc}   
            \toprule   
            \multicolumn{2}{c}{\multirow{2}[2]{*}{Method}} & \multicolumn{3}{c}{FIB25} & \multicolumn{3}{c}{Cremi} \\   
            \cmidrule(lr){3-5}\cmidrule(lr){6-8}
            \multicolumn{2}{c}{} & x2    & x4    & x8    & x2    & x4    & x8 \\
            \midrule    \multirow{2}[2]{*}{Interplation} & PSNR  & 33.21 &30.29 &29.19 & 31.44 &29.34 & 28.27  \\   
            & SSIM  &0.854 &0.722 &0.538 & 0.782 &0.574 & 0.451  \\
            \multirow{2}[2]{*}{$\dag$3DSRUNet} & PSNR  & \textbf{33.84} &32.31 &30.97 &\textbf{32.04} &31.12 &\textbf{30.28} \\     
            & SSIM  & 0.877 &0.824 &0.741 &\textbf{0.820} &0.761 &0.719 \\ 
            \multirow{2}[2]{*}{CycleGAN-IR} & PSNR  &33.54 &31.77  &29.94 &  31.71  &30.47 &29.04  \\        
            & SSIM  &0.869 &0.798  &0.640 &  0.794 &0.721 &0.560  \\
            \multirow{2}[2]{*}{\textbf{DiffuseIR (ours)}} & PSNR  &33.81 & \textbf{32.37} &\textbf{31.09} & 31.97 &\textbf{31.24} & 30.24  \\     
            & SSIM  &\textbf{0.881} &\textbf{0.832} &\textbf{0.774} & 0.819 &\textbf{0.783} & \textbf{0.726}  \\
            \bottomrule   
        \end{tabular}% 
        \begin{tablenotes}
            \item $\dag$ Supervised method.
        \end{tablenotes}
    %}
\end{table*}%

\subsubsection{Quantitative and Visual Evaluation.}
\begin{figure}[htb]
\centering
\includegraphics[width=1.0\linewidth]{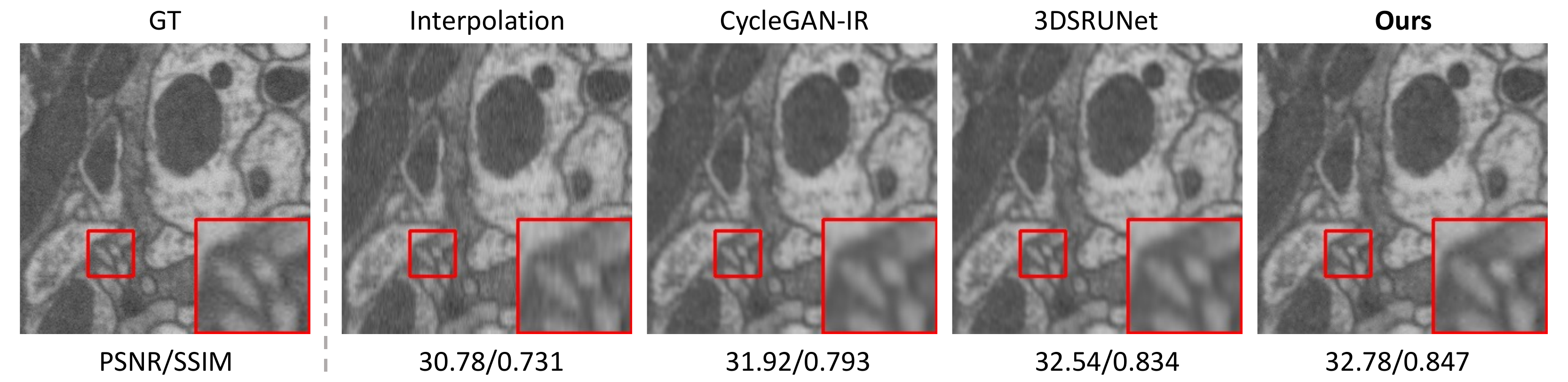}
\caption{\label{fig:2}\textbf{Visual comparisons on FIB-25 dataset ($\alpha=4$).} DiffuseIR can generate competitive results compared to supervised methods, and the results appear more visually realistic.}   
\label{method_framework}
\end{figure}

To evaluate the effectiveness of our method, we compared DiffuseIR with SoTA methods and presented the quantitative results in Tab. \ref{main_table1}. We use cubic interpolation as a basic comparison. 3DSRUNet \cite{LarissaHeinrich2017DeepLF} is a seminal isotropic reconstruction method based on deep learning, which requires high-resolution and low-resolution pairs as ground truth for supervised training. CycleGAN-IR \cite{ShiyuDeng2020IsotropicRO} proposed an unsupervised approach using a CycleGAN \cite{JunYanZhu2017UnpairedIT} architecture, learning from unpaired axial and lateral slices. 
It is worth noting that these methods train specialized models based on a fixed anisotropic spatial setting. In addition, they need to be retrained when facing different anisotropic factors $\alpha$. Various anisotropic factors $\alpha$ are shown in Tab. \ref{main_table1}.
Despite the model $\theta$ is trained solely for denoising tasks and having no exposure to axial slices during training, DiffuseIR outperforms unsupervised baselines and is even competitive with the supervised method \cite{LarissaHeinrich2017DeepLF}. As shown in Fig. \ref{fig:2}, using our proposed refine-in-loop strategy, the results produced by DiffuseIR exhibit more significant visual similarity to the Ground Truth compared to other methods, which may be more prone to causing distortion and blurriness of some details.
Notably, the versatility afforded by SSCS allows DiffuseIR to achieve excellent results using only one model, even under different isotropic resolution settings. 
% This indicates that DiffuseIR overcomes the issue of generalization to some extent in practical scenarios, as users no longer need to retrain the model after modifying imaging settings.
This indicates that DiffuseIR overcomes the issue of generalization to some extent in practical scenarios, as users no longer need to retrain the model after modifying imaging settings.

% \label{main_table1}
% \centering
% \setlength\tabcolsep{8pt}%调列距
% \begin{tabular}{|l|ccc|ccc|}
% \hline
%  Method& &FIB25 & & &Cremi & \\ \hline
% & x2 & x4 & x8 & x2 & x4 & x8\\ \hline
%  Interplation &33.21 &30.29 &29.19 & 31.84 &29.21 & 27.17  \\
% &0.854 &0.722 &0.518 & 0.739 &0.654 & 0.525  \\ \hline
%  3DSRUNet\cite{LarissaHeinrich2017DeepLF} & \textbf{33.84} &32.51 &31.61 &31.65 &31.42 &30.69 \\
% & 0.877 &0.811 &0.741 &0.834 &0.741 &0.844 \\ \hline
% CycleGAN-IR\cite{ShiyuDeng2020IsotropicRO} &33.62 &31.47  &29.80 &  30.41  &29.93 &29.04  \\
% &0.869 &0.775  &0.560 &  0.726 &0.761 &0.610  \\ \hline
% Ours &33.81 & 32.97 &31.99 & 32.31 &31.44 & 30.48  \\ 
% &\textbf{0.881} & 0.832 &0.764 & 0.816 &0.773 & 0.836  \\ \hline
% \end{tabular}
% \end{table*}

%v2:

 \begin{figure}[H]
\centering
\includegraphics[width=\linewidth]{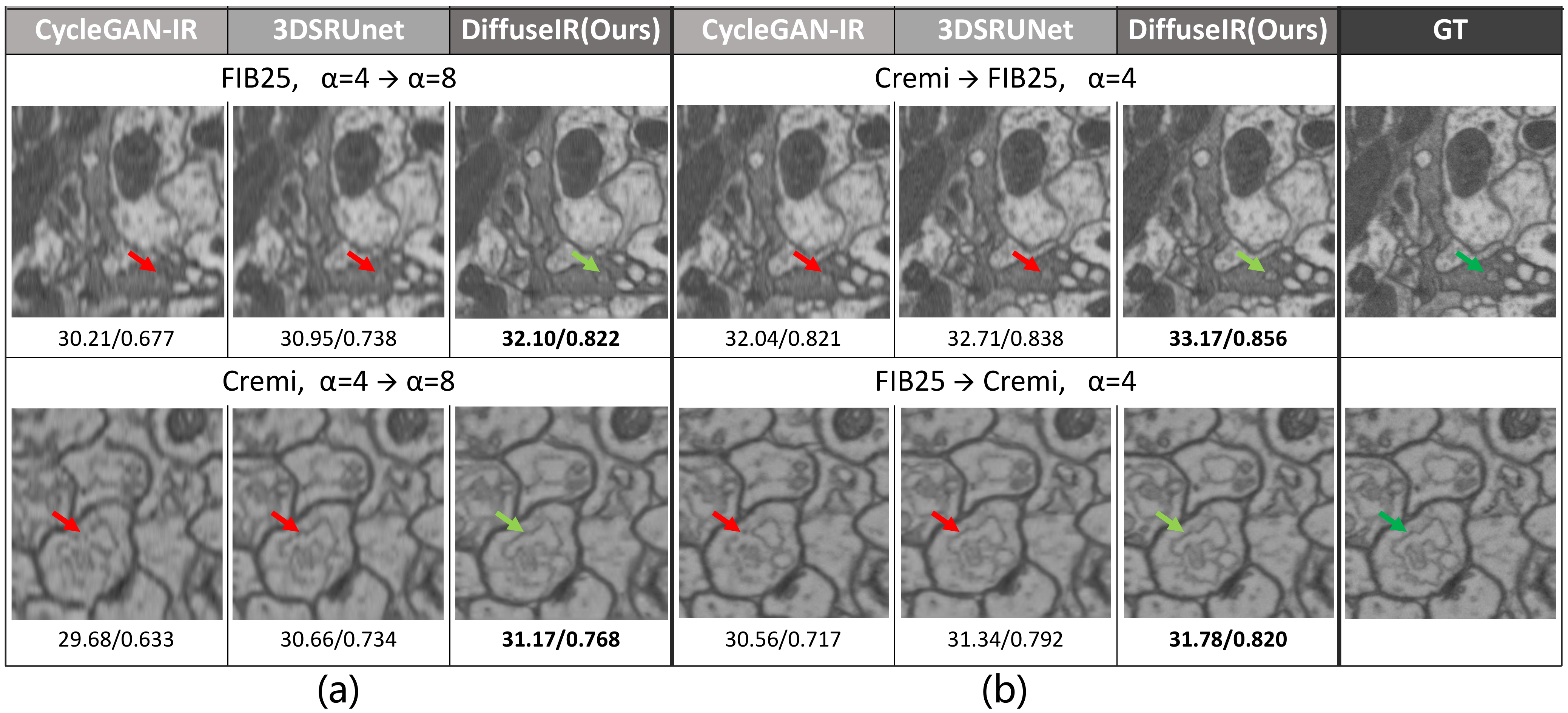}
\caption{\label{fig:3}\textbf{Analysis on robustness.} (a) Test on unseen anisotropic factor $\alpha$.
% (e.g., train on $\alpha=4$, test on $\alpha=8$).
(b) Test on different datasets with domain shifts (e.g., train on FIB25, test on Cremi). Our method is robust against various anisotropic factors and domain shifts between two datasets.}
\end{figure}

\subsubsection{Further analysis on robustness.}
%V2:
We examined the robustness of our model to variations in both Z-axis resolutions and domain shifts. Specifically, we investigated the following:
\textbf{(a) Robustness to unseen anisotropic spatial factors.} The algorithm may encounter unseen anisotropic resolution due to the need for different imaging settings in practical applications.
To assess the model's robustness to unseen anisotropic factors, we evaluated the model trained with the anisotropic factor $\alpha=4$. Then we do inference under the scenario of anisotropic factor $\alpha=8$. For those methods with a fixed super-resolution factor, we use cubic interpolation to upsample the reconstructed result by 2x along the axis.
\textbf{(b) Robustness to the domain shifts.} When encountering unseen data in the real world, domain shifts often exist, such as differences in biological structure features and physical resolution, which can impact the model's performance \cite{GabrielaCsurka2017DomainAF,HaoGuan2021DomainAF}. To evaluate the model's ability to handle those domain shifts, we trained our model on one dataset and tested it on another dataset.

%(FIB25$\rightarrow$Cremi and Cremi$\rightarrow$FIB25).

% \textbf{Analysis}: As shown in Fig. \ref{fig:3}, DiffuseIR exhibits significantly stronger robustness compared to other methods. For scenario \textbf{(a)}, other methods are trained on specific anisotropic factors for the super-resolution task of axial LR to lateral HR, which can result in model fragility when applied to different resolution settings during testing. In contrast, DiffuseIR directly learns the HR distribution at lateral through a generative task, which naturally applies to various axial resolutions. For scenario \textbf{(b)}, all methods show a decrease in performance, but DiffuseIR exhibits the smallest relative performance drop. We believe this is due to the stability provided by the multi-step generation of the diffusion model, combined with the sparse spatial constraints imposed by SSCS at each reverse-diffusion step, enabling safe generation of images that follow the distribution $p_{\theta}(X_{lat})$. In contrast, other methods generate results in one step using CNN networks, which can result in catastrophic and incomprehensible predictions when faced with images with significant domain gaps.

\textbf{Analysis}: As shown in Fig. \ref{fig:3}, DiffuseIR shows greater robustness than other methods. In scenario \textbf{(a)}, other methods are trained on specific anisotropic factors for super-resolution of axial LR to lateral HR. This can result in model fragility during testing with unseen anisotropic resolutions. In contrast, DiffuseIR directly learns the universal structural distribution at lateral through generation task, applicable to various axial resolutions.
All methods exhibit decreased performance in scenario \textbf{(b)}. However, DiffuseIR shows a small performance degradation with the help of the multi-step generation of the diffusion model and sparse spatial constraints imposed by SSCS at each reverse-diffusion step.
Further, compared to the previous methods predicting the result by one step, DiffuseIR makes the generating process more robust and controllable by adding constraints at each step to prevent the model from being off-limit.
%enables the safe generation of images following distribution $p_{\theta}(X_{lat})$. Other methods directly generate results in one step using CNN networks, leading to catastrophic and incomprehensible predictions with significant domain gaps.

% For biological and medical data, the reliability of the model is crucial, and it is necessary to maintain the performance as much as possible when facing new shooting equipment or tissues. We conducted cross-dataset testing and tested the model trained in one dataset on another. In Table \ref{OOD}, our method has only slight performance degradation compared to Table \ref{main_table1}, and the performance degradation of baselines are more significant. It is worth mentioning that our method even surpasses the supervised baseline, which reflects good extraterritorial generalization. 

\begin{figure*}[t]
\centering
\includegraphics[width=1.0\linewidth]{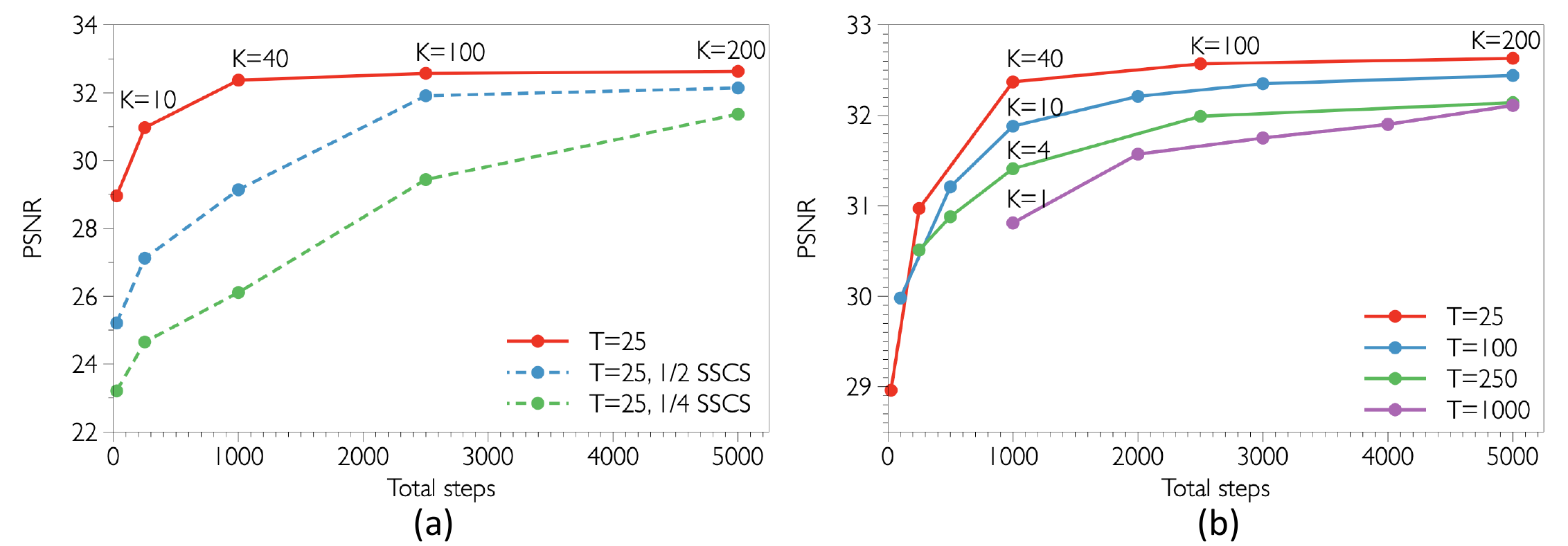}
\caption{\label{fig:4}\textbf{Ablation Study:} \textbf{(a)ablation on SSCS frequency.}Experimental results demonstrates the importance of SSCS. When reducing the frequency of SSCS usage, performance will severely decline. \textbf{(b)ablation on different refine-in-loop settings.} The results show that when the number of total steps is fixed, increase K will lead to higher PSNR.}
\end{figure*}

\subsubsection{Ablation Study.} 
\label{ablation}
We conducted extensive ablation experiments Fig. \ref{fig:4}. First, to demonstrate the effectiveness of SSCS, we use it only in partially alternate reverse-diffusion steps, such as 1/4 or 1/2 steps. As shown in Fig. \ref{fig:4} (a), increasing the frequency of SSCS significantly improves PSNR while bringing negligible additional computational costs. This indicates that SSCS have a vital effect on the model's performance.
Second, for the Refine-in-loop strategy, results show that keeping the total number of steps unchanged (reducing the number of time steps T while increasing the refine iterations K) can markedly improve performance. Fig. \ref{fig:4} (b) have the following settings: $T=\{25,100,250,1000\}$ with $K\{40,10,4,1\}$ to achieve a total of 5000 steps. The results show that the model performs best when $T=25$ and PSNR gradually increases with the increase of $K$. A balanced choice is $\{T=25, K=40\}$, which improves PSNR by 1.56dB compared to $\{T=1000, K=1\}$ without using the Refine-in-loop strategy.

% DiffuseIR一定程度上解决了当前方法在真实场景中的通用性问题，并有着鲁棒、可控的特点。然而，不可否认的是diffusion model会带来额外的更多的计算开销，需要长时间的离线处理。我们的Refine-in-loop策略发现减少diffusion steps并提高loop次数能够以更少的steps取得高性能，这一定程度上缓解了计算开销大的问题，用户可以自己在性能和速度之间选择参数进行平衡。最近，diffusion model的加速采样被视为重要的问题一直在被推进，我们相信在未来工作中DiffuseIR能够受益于社区中的进步并进一步减少计算时间。
% \textbf{[TO BE DELETED.]}DiffuseIR addresses the issue of generalization in practical scenarios to some extent and has robust and controllable features. However, diffusion models introduce additional computational overhead and require a long offline processing time. Our Refine-in-Loop strategy has shown that reducing the number of diffusion steps and increasing the loop count can achieve high performance with fewer steps, mitigating the issue of high computational costs and allowing users to balance parameters between performance and speed. Recently, accelerated sampling of diffusion models has been considered an important problem, and we believe that DiffuseIR will benefit from advances in the community and further reduce computational overhead in future work.

\section{Conclusion}
% HH：内容和摘要差不多，强调一下方法设计的合理性和意义，量化验证结果好，升华该研究在领域中的意义
We introduce DiffuseIR, an unsupervised method for isotropic reconstruction based on diffusion models. To the best of our knowledge, We are the first to introduce diffusion models to solve this problem. Our approach employs Sparse Spatial Condition Sampling (SSCS) and a Refine-in-loop strategy to generate results robustly and efficiently that can handle unseen anisotropic resolutions. We evaluate DiffuseIR on EM data.  
%DiffuseIR achieves SoTA performance, even being competitive with the supervised method. 
Experiments results show our methods achieve SoTA methods and yield comparable performance to supervised methods. 
Additionally, our approach offers a novel perspective for addressing Isotropic Reconstruction problems and has impressive robustness and generalization abilities.

% % HH：这里一般写老师的基金，本文被哪个基金支持
% \subsubsection{Acknowledgements} Please place your acknowledgments at
% the end of the paper, preceded by an unnumbered run-in heading (i.e.
% 3rd-level heading).

\bibliographystyle{splncs04}
\bibliography{mybibliography}

\end{document}

\title{Supplementary Material}
\maketitle        

\section{Supplementary visualization results}
\begin{figure}[!htb]
\centering
\includegraphics[width=1.0\linewidth]{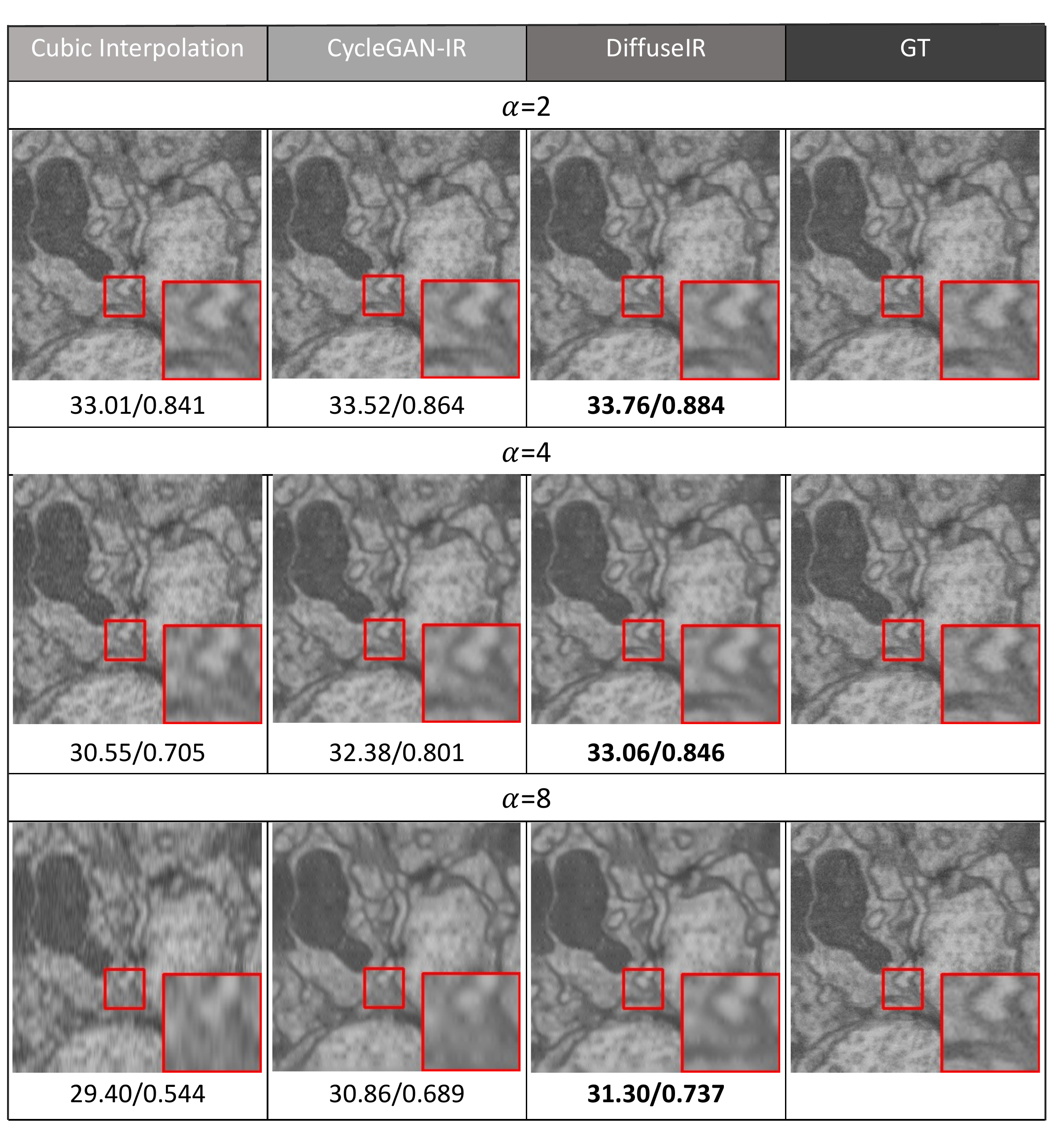}
\caption{\label{fig:2} Additional results of anisotropic resolution factor $\alpha=\{2,4,8\}$ on the FIB-25 dataset. Under the same conditions, DiffuseIR can generate images that are more realistic and have less artifacts.
}   
\label{method_framework}
\end{figure}

\section{Supplementary description of metrics}
We use PSNR and SSIM to evaluate the effectiveness of the reconstruction results between X and the corresponding ground truth image Y.

\subsubsection{PSNR}
\begin{gather*}
PSNR= 10 \log_{10}\left(\frac{{MAX}^2(X)}{MSE}\right) \\
{MSE} = \frac{1}{mn}\sum_{i=0}^{m-1}\sum_{j=0}^{n-1}[X(i,j) - Y(i,j)]^2
\end{gather*}

\subsubsection{SSIM}
\begin{gather*}
\text{SSIM}(x,y) = \frac{(2\mu_x\mu_y + \epsilon_1)(2\sigma_{xy} + \epsilon_2)}{(\mu_x^2 + \mu_y^2 + \epsilon_1)(\sigma_x^2 + \sigma_y^2 + \epsilon_2)}
\end{gather*}
where $x$ and $y$ denote the sliding windows of $X$ and $Y$, $\mu_x$ and $\mu_y$ are the mean values, $\sigma_x$ and $\sigma_y$ are the variance values, and $\sigma_{xy}$ is the covariance between $x$ and $y$. Also, $\epsilon$ is the constant used to avoid an undefined quotient.

\section{Supplementary description of DDIM}
As mentioned in the paper, DDIM \cite{JiamingSong2020DenoisingDI} can be used as an optional acceleration method to decrease the number of sampling steps by modifing the sampling process of \cite{JonathanHo2020DenoisingDP}. The diffusion process $q$ is derived as:
\begin{equation}
\begin{aligned}
$q(x_{t-1}|x_t,x_0)=N(\sqrt{\overline{\alpha}_{t-1}}x_0+
    \sqrt{1-\overline{\alpha}_{t-1}-\sigma^2} 
    \frac{x_t-\sqrt{\overline{\alpha}_{t}}x_0}{1-\overline{\alpha}_t},\sigma^2I)$
\end{aligned}
\end{equation}
The reverse-diffusion process $p$ is derived as:
\begin{equation}
\begin{aligned}
$x_{t-1} = \sqrt{\overline{\alpha}_{t-1}}(\frac{x_t-\sqrt{1-\overline{\alpha}_t}\epsilon_\theta(x_t,t)}{\sqrt{\overline{\alpha}_t}}) + \sqrt{1-\overline{\alpha}_{t-1}-\sigma_t^2}\epsilon_\theta(x_t,t) + \sigma_tz_t$ \;
        % N(x_{t-1};,\sigma_t^2I)$ \;
\end{aligned}
\end{equation}     
where $\epsilon_\theta$ is the pre-trained diffusion model which learns a distribution at lateral.
\bibliographystyle{splncs04}
\bibliography{mybibliography}